\documentclass[journal]{IEEEtran}

\usepackage{cite}

\ifCLASSINFOpdf
  \usepackage[pdftex]{graphicx}
  \graphicspath{ {Graphics/} } 
  \DeclareGraphicsExtensions{.pdf,.jpeg,.png}
\else
  \usepackage[dvips]{graphicx}
  \graphicspath{{../eps/}}
  \DeclareGraphicsExtensions{.eps}
\fi

\usepackage{amsmath}

\usepackage{algorithmic}

\usepackage{array}

\ifCLASSOPTIONcompsoc
 \usepackage[caption=false,font=normalsize,labelfont=sf,textfont=sf]{subfig}
\else
 \usepackage[caption=false,font=footnotesize]{subfig}
\fi
\usepackage{url}

\usepackage{amsfonts}
\newcolumntype{P}[1]{>{\centering\arraybackslash}p{#1}}
\newcolumntype{M}[1]{>{\centering\arraybackslash}m{#1}}
\usepackage{tabularx}
\usepackage{multirow}
\usepackage{siunitx}
\usepackage{footnote}
\usepackage{tablefootnote}
\usepackage{threeparttable}
\usepackage{caption}
\usepackage{booktabs}
\usepackage{threeparttable,booktabs}
\usepackage{subfig}

\begin{document}
%
\title{Voltage Estimation in Low-Voltage Distribution Grids with Distributed Energy Resources}

\author{Marija~Markovi\'{c},~\IEEEmembership{Graduate Student Member,~IEEE,}
        Amirhossein~Sajadi,~\IEEEmembership{Senior~Member,~IEEE,}
        Anthony~Florita,~\IEEEmembership{Member,~IEEE,}
        Robert~Cruickshank III,~\IEEEmembership{Senior~Member,~IEEE,}
        and~Bri-Mathias~Hodge,~\IEEEmembership{Senior~Member,~IEEE}
\thanks{M. Markovi\'{c},  A. Sajadi, and B.-M. Hodge are with the University of
Colorado Boulder, Boulder, CO, 80309 USA.}
\thanks{B.-M. Hodge and A. Florita are also with the National Renewable Energy Laboratory (NREL), 
Golden, CO 80401 USA.}
\thanks{R. Cruickshank III is with the Cable Television Laboratories, Louisville, CO 80027 USA and Robert Cruickshank Associates, Big Indian, NY 12410 USA}
}

\maketitle

\begin{abstract}
The present distribution grids generally have limited sensing capabilities and are therefore characterized by low observability. Improved observability is a prerequisite for increasing the hosting capacity of distributed energy resources such as solar photovoltaics (PV) in distribution grids. In this context, this paper presents learning-aided low-voltage estimation using untapped but readily available and widely distributed sensors from cable television (CATV) networks. The CATV sensors offer timely local voltage magnitude sensing with 5-minute resolution and can provide an order of magnitude more data on the state of a distribution system than currently deployed utility sensors. The proposed solution incorporates voltage readings from neighboring CATV sensors, taking into account spatio-temporal aspects of the observations, and estimates single-phase voltage magnitudes at all non-monitored buses using random forest. The effectiveness of the proposed approach was demonstrated using a 1572-bus feeder from the SMART-DS data set for two case studies – passive distribution feeder (without PV) and active distribution feeder (with PV). The analysis was conducted on simulated data, and the results show voltage estimates with a high degree of accuracy, even at extremely low percentages of observable nodes.
\end{abstract}


%
\IEEEpeerreviewmaketitle

\section{Introduction} \label{Intro}
%
%
%
\IEEEPARstart{H}{igh} penetration of variable generation (VG) and distributed energy resources (DERs) fundamentally alter the dynamics of distribution systems, leading to numerous challenges from the system's operation and reliability standpoints \cite{6248721}. Voltage rise due to distributed PV is a common issue i.e., maintaining voltages within the ANSI C84.1 operating tolerances will become more difficult with higher shares of PV due to limited observability as a result of insufficient availability of real-time measurements. Insufficiently accurate and untimely grid information stems in principle from (i) scarce sensor counts, and (ii) inadequate communications infrastructure at distribution level \cite{dehghanpour2018survey}. Accordingly, distribution system operators (DSOs) need to expand their system monitoring efforts to effectively operate the grid in response to emerging challenges. 

Over the past decade, a mass deployment of smart meters (SMs) has laid a promising basis for increasing distribution grid observability \cite{bhela2017enhancing}. In recent years, U.S. distribution utilities have deployed about 86.8 million SMs, of which about 88\% are installed at residential customers' premises \cite{noauthor_electric_2019}. Significant research has been conducted to leverage SM data, in particular for distribution system state estimation (DSSE) \cite{alimardani2014using, alimardani2015distribution, primadianto2016review, pau2016low, alzate2019distribution, kemal2020trade}. However, addressing the technical (e.g., time synchronization and slow sampling rates of SM readings \cite{sanchez2017observability}) and seemingly intractable non-technical issues (e.g., end-user privacy concerns) associated with SMs requires further research efforts and policy formulations. Extensive work has also been done on optimizing the placement of additional metering instruments by selecting strategic measuring points to increase network observability \cite{7741933, furlani2020machine}. Although the expansion of metering and communications infrastructure is a desirable solution to overcome existing observability constraints, it is unfortunately economically and time-wise impractical due to the immense number of distribution system nodes.
 
Recent efforts have been made to apply machine learning (ML) for non-conventional voltage estimation in distribution systems \cite{pertl2016voltage, furlani2020machine, 8913505}. For instance, \cite{pertl2016voltage} proposed Artificial Neural Networks algorithm as a low-voltage (LV) state estimator, which estimates voltage magnitudes without the knowledge on underlying physical model of the system. More recent work \cite{furlani2020machine} explored multiple ML regression algorithmic solutions, including ensemble of regressors, to predict distribution grid voltages, also without complete knowledge of the network topology. While in \cite{pertl2016voltage} only available measurements of already deployed SMs were assumed, in \cite{furlani2020machine}, the voltage prediction emerged from the premise of full availability of active and reactive powers obtained from SMs at load nodes. Several recent works have focused on addressing the need for DSSE implementation under low-observability conditions \cite{8930601, 8913505}. In \cite{8913505}, the Bayesian linear least squares estimator is used to forecast all non-measured voltages under limited sensing capabilities. Recent work \cite{8930601} has also developed a DSSE technique applicable in low-observability cases that offers additional flexibility as it is not restricted to specific measurements. 

Similar to recent works \cite{8930601, 8913505}, this paper aims to improve observability in LV distribution grids by enabling accurate voltage estimation of all non-monitored buses using a limited number of measurements. Voltage estimation is performed using random forest, a supervised ML algorithm, known for its high robustness and easy implementation \cite{breiman2001random}. The main application of the proffered solution is to support DSOs operational decision-making in low-observability conditions and thus aid the operation of LV distribution feeders challenged by the growing presence of VG and DERs. The framework presented herein leverages existing CATV voltage sensing through an inherently available secure, high-bandwidth, low-latency communications network, resulting in minimal or no additional capital expenditures. While this particular application employs CATV voltage measurements, it is not restrictive of the sensor type and can use voltage measurements available from any other sensors. However, the main advantage of CATV measurements over other utility sensor measurements is their fine (5-minutes) resolution and already established communications network that provides data transport at gigabit speeds and millisecond latencies. As most consumers in the LV distribution grids are single-phase consumers \cite{6248721}, the estimation of single-phase voltages is formally presented in this work. However, the proposed approach can easily be extended to multi-phase voltage estimation through the multi-output regression capability of random forest. The underlying methodology and motivation come from the timely voltage sensing offered by CATV sensors, not previously addressed in literature. Note that the data used are not real-world data, but simulated case studies, representative of actual CATV measurements. The contributions of this paper are twofold:

\begin{itemize}
  \item A novel learning-aided voltage estimation framework with the aim of enabling a more realistic insight into current sparsely monitored LV distribution grids is presented. The accuracy of voltage estimates is achieved across different observability operating modes, including low-observability, where traditional least-squares state estimation formulations are not applicable.  
  \item The proposed framework obviates the need for SM data. The prediction model relies solely on available voltage measurements from existing CATV sensors, however, not necessarily limited to the CATV sensors.
\end{itemize}

The rest of the paper is organized as follows: Section \ref{Motivation} outlines sensor potentials from CATV broadband networks, from which the motivation for this paper originated. The proposed voltage estimation framework is presented in Section \ref{Observability Enhancement Framework}. Section \ref{Case Studies} details the data sets for the case studies examined. Section \ref{Results} presents the voltage prediction results from applying the framework to the case study data. Section \ref{Conclusion} concludes the work and provides suggestions for future research directions.

\section{CATV Sensing Potential} \label{Motivation}

While distributed solar PV and other DERs have the potential to transform power system operation, they also escalate system complexity and uncertainty. Accommodation of these emerging technologies requires a low-cost and scalable communications infrastructure to increase the situational awareness and mitigate the operational challenges associated with them.

\begin{figure}[!t]
\centering
\includegraphics[width=\columnwidth]{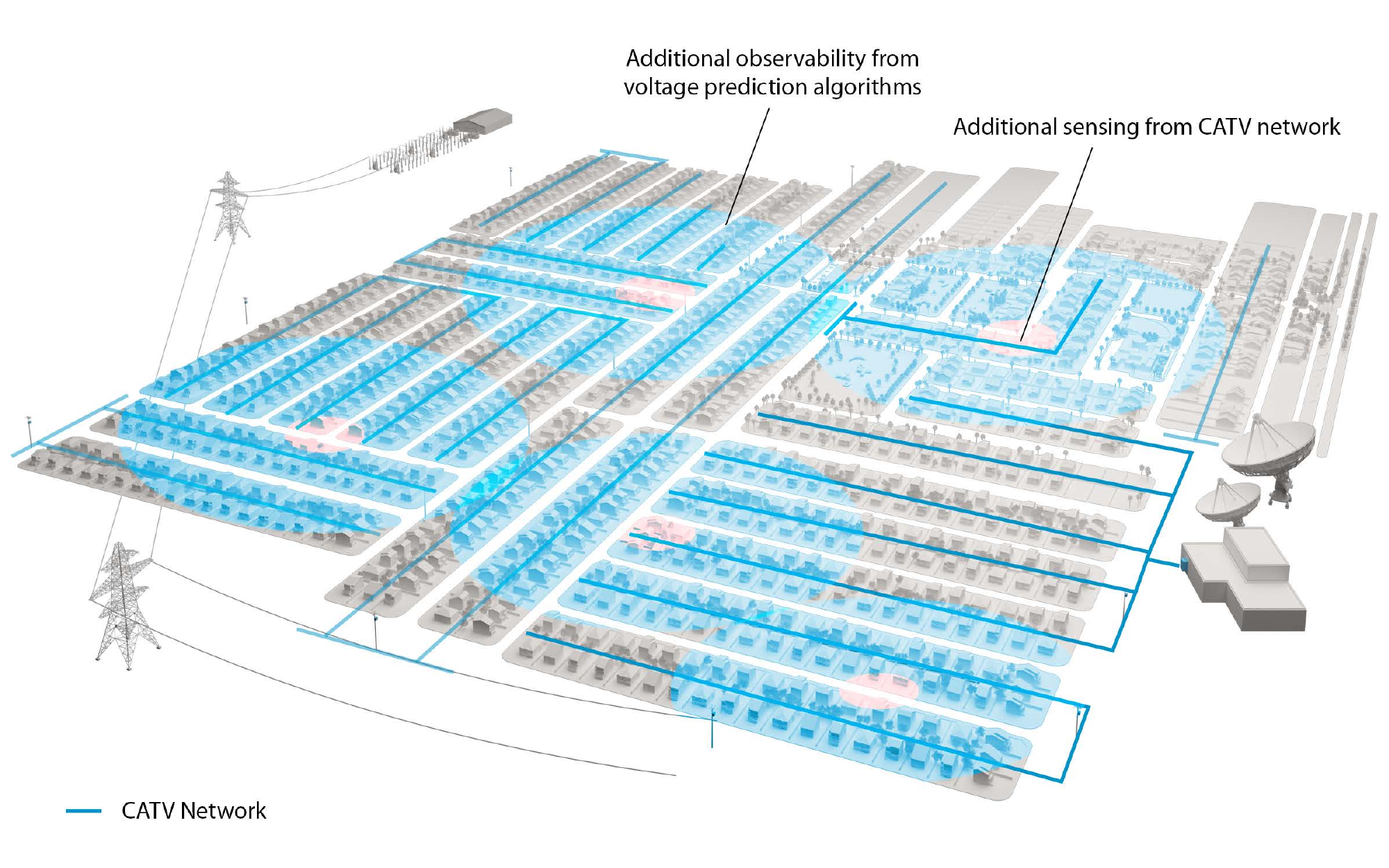}
\caption{Synergy of CATV and electric distribution networks, with the aim of leveraging voltage observations from CATV networks' sensors.}
\label{Connected Neighborhood}
\end{figure}

Fortunately, broadband cable networks, so-called CATV networks are widely distributed across the United States, with approximately 800,000 power sensors in U.S. and millions of sensors worldwide. CATV networks are physically separated from the electric grid and have sensors already deployed which reach overlay existing distribution grids, mapping to approximately 95\% of residential and commercial utility customers (Figure~\ref{Connected Neighborhood}). Therefore, the two networks, CATV and electric distribution, can be coupled and use the CATV communications network to provide timely measurements via sensors in battery-backed power supplies. Each sensor, typically connected to the secondary side of split-phase transformers, covers a geographic area of approximately 100-200 homes and measures the as-delivered secondary distribution grid voltage. Sensor readings have a sampling period less than 15-minutes, typically 5-minutes, and include input voltage and inverter status (i.e., online vs. standby operation indicating grid power loss). In the context of power systems, these data have not been utilized for power system management functions to date.

The CATV power supply sensor provides approximately 1 measurement per small group of residential or commercial customers, which has the potential to provide an order of magnitude more data on the state of a distribution system than the currently deployed utility sensors. Even where distribution utilities have employed AMI, voltage sensing of all deployed SMs is limited to 240-V measurements with a sampling frequency of 15-minutes, at best. This means that the data gathered from the broadband network power sensors already has a significantly higher resolution than current SMs data. The main advantage of CATV sensors over AMI is the existence of a high-bandwidth, low-latency, secure communications network that transmits data in compliance with the Data over Cable Service Interface Specification (DOCSIS), an international standard that has demonstrated resilient cyber-security against incidents \cite{Fischer2020}. The streamed broadband sensor voltage observations will add significant value given the existing secure communications network inherently available for their utilization. In other words, the CATV solution removes cost and construction time barriers for installation of distribution grid sensing and communications infrastructure.

Although a distance-based correlation between the local distribution grid and CATV sensor placements does not match one-to-one, it can be addressed with algorithmic solutions, such as cluster analysis. The untapped connection between these two large infrastructure systems provides a fundamental basis for the idea behind the proposed methodology described in section \ref{Observability Enhancement Framework}.

\section{LV Observability Enhancement Framework} \label{Observability Enhancement Framework}

\begin{figure}[!t]
\centering
\includegraphics[width=0.85\columnwidth]{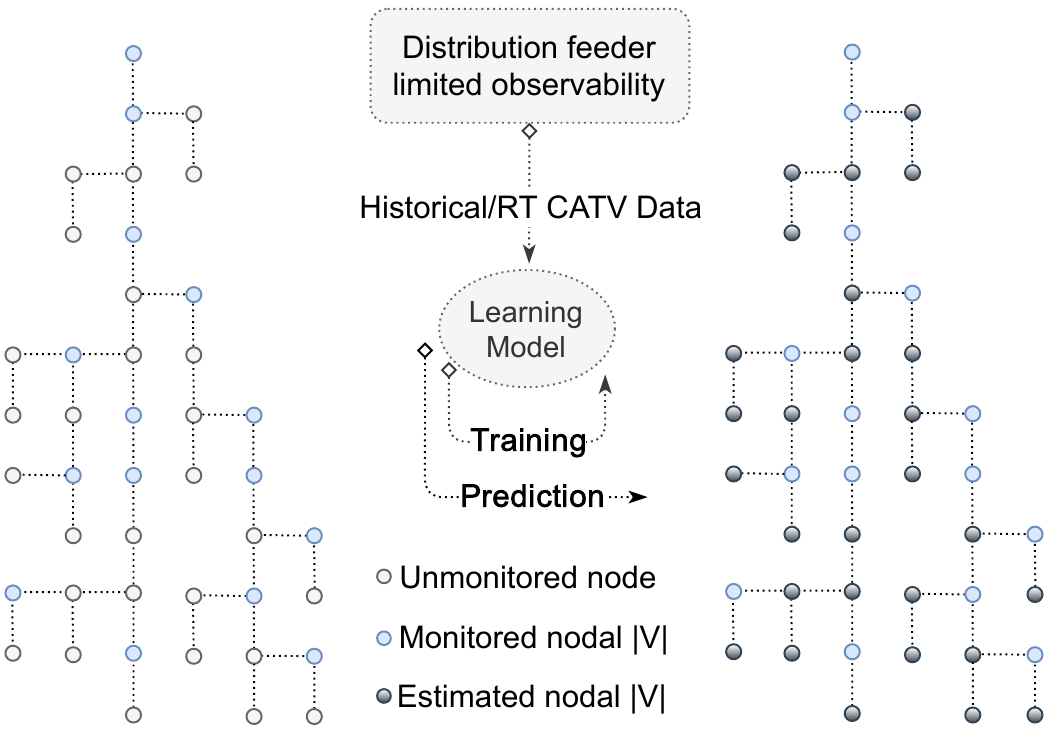}
\caption{Conceptual diagram of the proposed learning-aided voltage prediction framework.}
\label{framework}
\end{figure}

A novel framework for enhancing observability in LV distribution grids is proposed as an enabling solution to utilize the potential that CATV networks offer for wide-area measurement in distribution systems. The proposed framework learns numerical regression models to predict unmonitored nodal voltage magnitudes based on historical and/or real-time data available from CATV sensors. Figure~\ref{framework} depicts a concept of the proposed framework, which has two fundamental modes of operation. The first operation mode is \textit{training} which is performed to learn the regression prediction models. The second operation mode is a \textit{prediction} of voltage magnitudes at unmonitored nodes. After a completion of the prediction mode, the associated performance indicators are used to evaluate the prediction accuracy. The proffered framework can estimate voltages using both historical CATV measurements and real-time measurements. 

The learning model is considered a nonlinear function that maps the voltages measured by CATV sensors into the voltages of neighboring LV grid nodes, taking into account the geospatial node coordinates. The mapping is obtained by random forest, briefly presented below. In this paper, two random forest algorithmic formulations were used to implement the prediction models. It should be noted that the term \textit{random forest} is a broader term used to denote any class of tree-structured ensemble algorithms. In the rest of the paper, the method \textit{random forest} refers specifically to the Random Forest (RF) implementation \cite{breiman2001random}, a very popular and widely used ML algorithm, based on the Classification and Regression Tree (CART\cite{breiman1984classification}) and Bootstrap Aggregating (\textit{Bagging} \cite{breiman2001random}) algorithms. Another \textit{random forest} variant used is tree-based \textit{Bagging} regression.

\subsection{Random Forest}
Random forest represents a family of tree-based ensemble methods, applicable to both supervised classification and regression tasks. In the case of classification, the output is a categorical label (class), while for regression problems the output is numerical (continuous). Unlike most supervised ML algorithms, random forest does not require normalization of input data, nor significant hyper-parameters tuning \cite{breiman2001random}, which is an advantage in terms of computational and memory consumption. The term \textit{training sample} henceforth denotes the observations used to build a model, \textit{n} refers to the size of the training sample (the number of observations), whereas \textit{p} is the number of features. 

Starting from the training sample \(\mathcal{T}\), consisting of \textit{n} pairs of target (output) variables $(y_{i})$ and \textit{p}-dimensional predictor (input) vectors $(\mathbf{x}_{i})$ ($\mathbf{x}_{i} \in \mathbb{R}^p ,y_{i} \in \mathbb{R}, i=1,...,n$), a total of \(\mathcal{B}\in \mathbb{N}\) bootstrapped samples are created via the Bagging algorithm \cite{breiman2001random}. The term \( \widetilde{\mathcal{T} }^b\) denotes bootstrap samples, each size \textit{n}, drawn at random but with replacement from the original training sample \(\mathcal{T}\) $(b=1,...,\mathcal{B})$. Each regression tree $T^b$, which constructs non-linear functions mapping from $\mathbf{x}_{i}$ to $y_{i}$, is then independently grown on a separate \( \widetilde{\mathcal{T} }^b\) \cite{breiman2001random}. To avoid a high correlation between regression trees, a random subset of \textit{q} features is selected without replacement from the total of \textit{p} at each decision split, where a value of $q=\frac{1}{3}p$ is proposed in the case of regression in \cite{hastie2009elements}. A splitting procedure is next performed based on the best splitting node candidate among the selected \textit{q}, afterwards the node is split into two daughter nodes. The quality of a split is measured by the mean squared error criteria \cite{pedregosa2011scikit}. This procedure is repeated recursively for each node independently, where each $T^b$ is grown to the largest extent possible (maximum depth), without pruning \cite{breiman2001random}. A final RF prediction $\widehat{f}^{rf}$ is obtained by averaging the predictions made by each regression tree in the ensemble (\ref{eq1}). 

\begin{equation}\label{eq1}
\widehat{f}^{rf}(\mathbf{x}) = \frac{1}{\mathcal{B}}\sum_{b=1}^{\mathcal{B}}T^b(\mathbf{x})
\end{equation}

In this paper, the optimal split is found in two ways: over a subset of features \textit{q}, but also over a whole set of input features \textit{p}. The former was originally suggested in \cite{breiman2001random} although the value of \textit{q} was not explicitly specified, whereas the latter is more recently suggested in \cite{geurts2006extremely} (note that when all predictor variables are used as split candidates on each node, then RF is no longer used, but, instead, the Bagging Regression Trees (BRT) algorithm, which is a special case when $q=p$). 

\section{Case Studies} \label{Case Studies}

To demonstrate the effectiveness of the proposed framework, two distinct operating scenarios are considered: (1) a passive distribution feeder, and (2) an active distribution feeder with distributed residential PV further described below.

\subsection{SMART-DS Synthetic Data Set}
The case studies are developed on a single feeder in the San Francisco Bay Area, taken from the SMART-DS synthetic data set \cite{krishnan2017smart, postigo2020phase}, with relevant load details presented in Table~\ref{table1}. The loads are modeled with different annual consumption characteristics, including commercial and residential time-series load profiles, available from SMART-DS. These time-series loads are based on 2012 and have 15-minutes resolution. The total number of potential sensor points corresponding to individual buses is 967 and 112 at the nominal system voltages of 208-V and 480-V, respectively.

\subsection{PV Model}
The second case study considered here involves distributed PV at residential load nodes. The PV systems are modeled with time-synchronous weather data from the same year as the load data. Solar irradiance and other relevant weather data were obtained from the NREL National Solar Radiation Database (NSRDB) \cite{sengupta2018national}, acquired from ten different sites located in San Francisco Bay Area to match the feeder location. It should be noted that the solar irradiance from the NSRDB database is of half-hourly resolution. To ensure time synchronization between load and solar profiles, all NSRDB weather data are taken from 2012 and downscaled to 15-minute resolution. The preprocessing is accomplished using a built-in macro of the open-source software System Advisor Model (SAM), which performs linear interpolation \cite{blair2014system}. Next, the PVWatts model in SAM is used to convert solar irradiance into the AC power output needed to build the PV model in the Open Distribution System Simulator (OpenDSS) software. The conversion is performed based on the following assumptions: (i) the average size of residential rooftop systems in California is 7kW \cite{fu_us_2018} (ii) DC to AC conversion ratio of the inverter is 1.2, and (iii) the inverter efficiency is 96\%. The last two assumptions are the default settings of the PVWatts model \cite{blair2014system}. All PV models implemented in OpenDSS are equally-sized rooftop residential PV systems connected to the laterals, and assigned randomly to residential load buses. The PV systems are modeled to operate under the unity power factor, according to the 1547-2013 standard \cite{6748837} (voltage regulation through reactive power control is not supported). Larger PV systems, usually connected to the MV, were not considered in this work.

\subsection{Open Distribution System Simulator}

The Python COM interface with OpenDSS was used to perform quasi-static time-series (QSTS) simulations on the 1572-bus feeder. Bus voltages were extracted iteratively at each simulation time step and separated based on the base voltage level for further processing. The resultant nodal voltage time-series sets enabled the establishment of "ground-truth" (benchmark) data against which the results obtained by prediction models can be compared. It should be noted that in this paper, \textit{node} and \textit{bus} are used interchangeably and denote the same term, whereas in the case of OpenDSS these two terms are distinguished.

\begin{table}[t!]
\caption{Feeder Load Details}
\label{table1}
\centering
    \begin{tabularx}{\columnwidth}{ccc}
Feeder & \begin{tabular}[c]{@{}c@{}}Rated Total Load\\ {[}MW{]} \end{tabular} & 
\begin{tabular}[c]{@{}c@{}} No. Residential/\\ Commercial Loads\end{tabular} \\ \hline
 \hline \rule{0pt}{4ex} 
p1uhs0\_1247--p1udt1470 & 21.62 & 594/149 
    \end{tabularx}
\end{table}

\begin{figure}
  \centering
  \subfloat[a][Case study I]{\includegraphics[width=0.95\columnwidth]{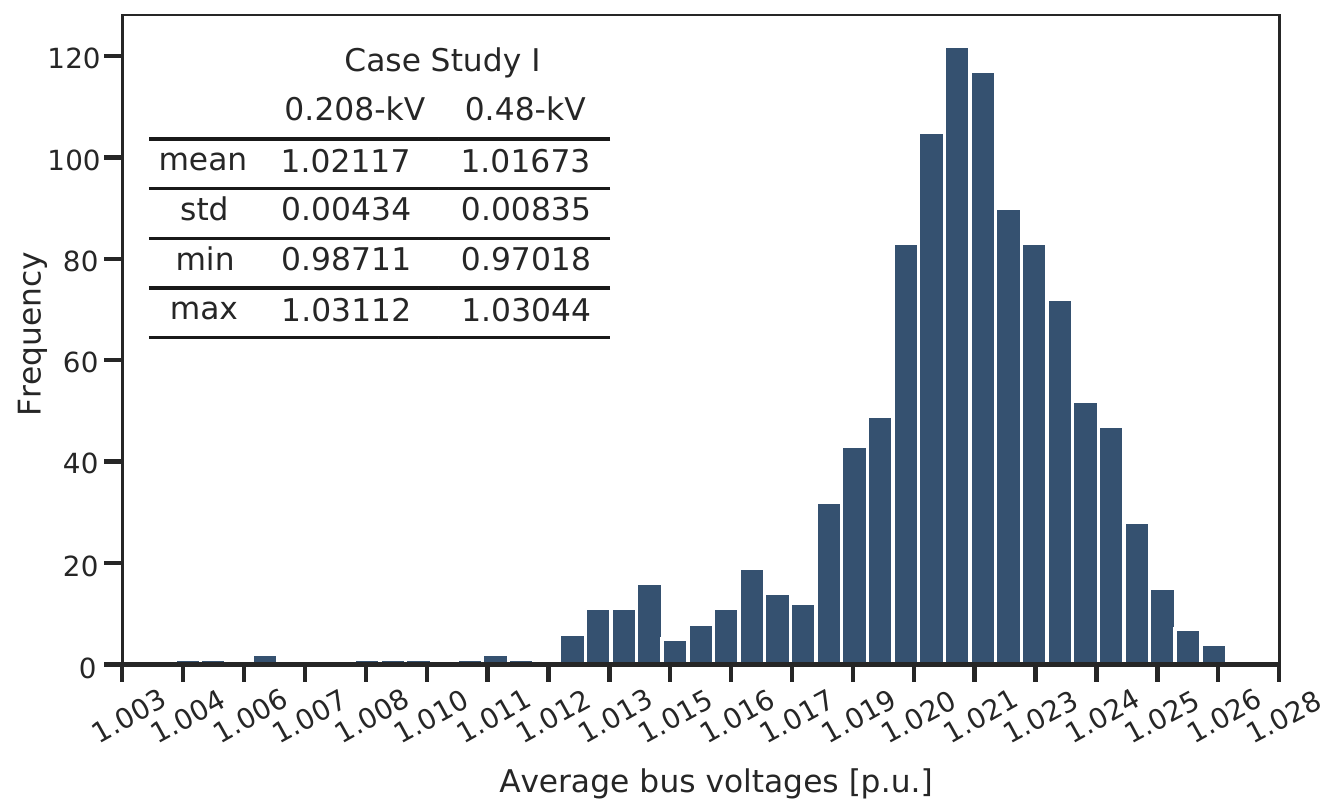} \label{fig:a}} \\
  \subfloat[b][Case study II]{\includegraphics[width=0.95\columnwidth]{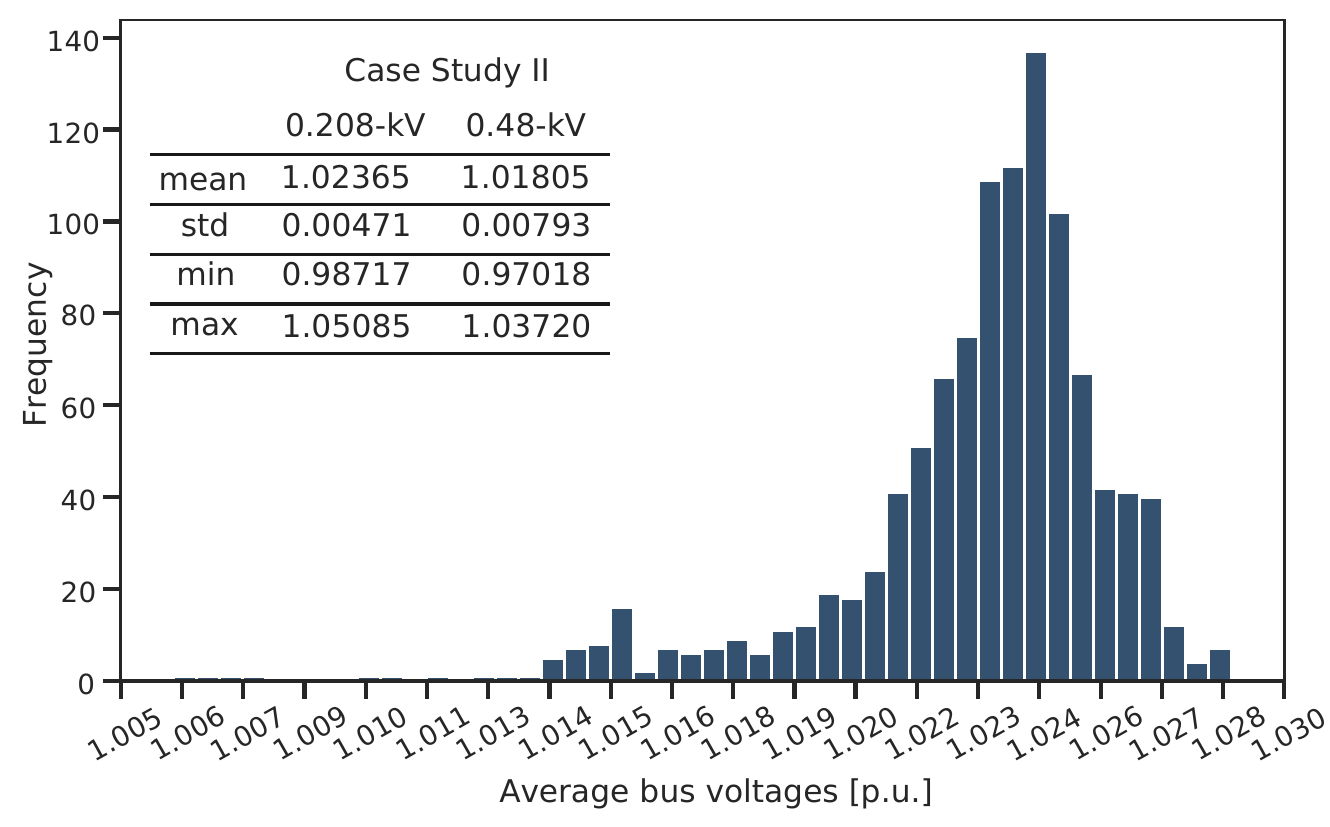} \label{fig:b}}
  \caption{A frequency distribution of average p.u. bus voltages with descriptive statistics of LV time-series profiles.} \label{Voltages Distribution}
\end{figure}

The first QSTS simulation was run on the original SMART-DS feeder (without PV) and the obtained benchmark data represent the first case study. The statistical distribution of bus voltages for the first case study, passive distribution feeder (normal operating conditions, $\pm5\% V_{n}$), is depicted in Fig.~\ref{fig:a}. The second QSTS simulation was performed on the same feeder modified to accommodate residential PV systems. In order to model the relatively high PV penetration in residential loads, 85\% of the buses (632 in total) were randomly selected and the PV systems were assigned to the corresponding loads. Specifically, 582 PV were added to 208-V buses, whereas the remaining 47 PV were added to 480-V buses, with a total installed PV capacity of 4.4 MW. The statistical distribution of bus voltages for the second case study is shown in Fig.~\ref{fig:b}. The voltages shown shown in Fig.~\ref{Voltages Distribution} correspond to all LV system bus voltages and serve to provide primarily descriptive insight into the base case data, but also as a reference for comparison with the voltages of the second case study with PV included. By comparing the voltage distributions for both cases, the shift of the LV voltage profile distribution to the right is clearly visible due to the addition of residential PV. Voltage fluctuations are not as large as they would be if higher capacity PV systems were added to the MV level. However, the amount of added PV is intentionally targeted for conditions close to the upper ANSI voltage limits. The most noticeable changes in the voltage profiles occur at 208-V, where most PV systems are distributed, which can be seen from the tables shown in Fig.~\ref{Voltages Distribution}. These tables include descriptive statistics of voltage time-series profiles for both nominal LV.

\begin{table}[t!]
\centering
\begin{threeparttable}
\caption{Case Studies Description}
\label{table1b}
\begin{tabular}{ ccccc }
\toprule
\begin{tabular}[c]{@{}c@{}}Case \\ Study\end{tabular} & Description & 
\begin{tabular}[c]{@{}c@{}}Voltage\\ {[}kV{]}\end{tabular} & \begin{tabular}[c]{@{}c@{}}Observability\\ Cases\end{tabular} & \begin{tabular}[c]{@{}c@{}}Random\\ Samplings\end{tabular}  \\
\toprule 
\multirow{6}{*}{I} & \multirow{6}{*}{\shortstack{Base\\ Network}} & \multirow{3}{*}{0.48} & Low & \multirow{3}{*}{3} \\
    		       &                               &                                      & Moderate & \\    		          
    	           &                               &                                      & High & \\
    	           &                               & \multirow{3}{*}{0.208} & Low & \multirow{3}{*}{3} \\
    		       &                               &                        & Moderate & \\    		          
    	           &                               &                        & High & \\
\midrule
\multirow{3}{*}{II} & \multirow{3}{*}{\shortstack{Base Network\\ + PV}} & \multirow{3}{*}{0.208\tnote{a}} & Low & \multirow{3}{*}{3} \\
    		        &                                    &                                       & Moderate & \\    		          
    	            &                                    &                                       & High & \\
\bottomrule
\end{tabular}
      \begin{tablenotes} \footnotesize
         \item[a] The second case study also includes 0.48-kV, but the results presented in the paper correspond to only 0.208-kV.
      \end{tablenotes}
\end{threeparttable}
\end{table}

\subsection{Observability Analysis}
In this paper, various levels of observability are studied. To this end, case studies that represent different degrees of observability are developed. To generate the necessary data sets for such observability analysis, the benchmark data of each main case study are used as a starting point. These data sets are created by sampling observable nodes. It should be noted that the availability/unavailability of voltage measurements at the respective buses was determined completely at random. That is, we randomly selected sets of buses (ranging between 1-80\% of buses) at which all potentially available voltages are measured via CATV sensors, while all other buses were considered unobservable. To minimize the likelihood of accidentally selecting nodes that could potentially lead to the best-case scenario for observability analysis, all data sets corresponding to the partially observable feeder were generated through three different random samplings and tested accordingly. Random sampling of observable nodes from benchmark data resulted in the generation of a total of 288 data sets (Table~\ref{table1b}). The degree of observability is modelled by different percentages of the potentially available nodal voltages, ranging from 1\% to 80\%, with an increment of 1\% for the first 10 data sets (corresponding to the low-observability), and with an increment of 5\% for the remaining 14 data sets (corresponding to cases of moderate- to high-observability). This allows testing of the proposed prediction models on a partially observable feeder, with different levels of observability, but still comparing the results at all locations with the fully observable system as a benchmark. 

\subsection{Model Implementation Specifics}
A total of 100 regression trees are used to build an ensemble, for both \textit{random forest} algorithms. The size of the training sample is 35,040 (data sets analyzed are annual time-series data with 15-minute granularity). Voltage magnitudes at unmonitored buses are target variables $y_{i}$; $\mathbf{x}_{i}$ is 6-dimensional and includes available voltage measurements, with corresponding features: nominal system voltage; bus label/number; bus geospatial (\textit{x} and \textit{y}) coordinates; and time instance at which the observations are preserved. The percentage of data used to train and test prediction models is directly determined by the percentage of monitored and unmonitored nodes, respectively.

\begin{table}[t!]
	\centering
    \begin{threeparttable}
    	\caption{Case Study I Average Metrics}
    	\label{tab:I}
    	\begin{tabular}{ c c c c }
    	    \toprule
            \begin{tabular}[c]{@{}c@{}}Voltage\\ {[}V{]}\end{tabular} & Model & \begin{tabular}[c]{@{}c@{}}RMSE\\ {[}\num{e-3} p.u.{]}\end{tabular} & \begin{tabular}[c]{@{}c@{}}MAE\\ {[}\num{e-3} p.u{]}\end{tabular} \\
    		\toprule 
    	    \multirow{3}{*}{208} & Random Forest & 2.008 & 1.393 \\
    		                    & Bagging Regression Trees & 2.283 & 1.558 \\    		          
    		                    & Linear Regression & 4.152 & 3.400	\\
    		\midrule
    		\multirow{3}{*}{480} & Random Forest & 5.716 & 4.201	\\
    		                    & Bagging Regression Trees & 6.166 & 4.348	\\    		          
    		                    & Linear Regression & 8.592\tnote{a} & 6.909\tnote{b} \\
    		\bottomrule
    	\end{tabular}
       \begin{tablenotes} \footnotesize
         \item[a] Excluded the two least observability cases metric values; with these two included, the average RMSE is \num{4.014e+7}.
         \item[b] Excluded the two least observability cases metric values; with these two included, the average MAE is \num{3.266e+7}. 
       \end{tablenotes}
    \end{threeparttable}
\end{table}

\begin{table}[t!]
	\centering
    \begin{threeparttable}
    	\caption{Case Study II Average Metrics}
    	\label{tab:II}
    	\begin{tabular}{ c c c c }
    	    \toprule
            \begin{tabular}[c]{@{}c@{}}Voltage\\ {[}V{]}\end{tabular} & Model & \begin{tabular}[c]{@{}c@{}}RMSE\\ {[}\num{e-3} p.u.{]}\end{tabular} & \begin{tabular}[c]{@{}c@{}}MAE\\ {[}\num{e-3} p.u{]}\end{tabular} \\
    		\toprule 
    	    \multirow{3}{*}{208} & Random Forest & 2.243 & 1.549 \\
    		                    & Bagging Regression Trees & 2.473 & 1.672 \\    		          
    		                    & Linear Regression & 4.583 & 3.436	\\
    		\bottomrule
    	\end{tabular}
    \end{threeparttable}
\end{table}

\subsection{Evaluation Metrics}
Two most common regression metrics are used to evaluate model performance; root mean square error (RMSE) and mean absolute error (MAE) given with expressions \ref{eq2}-\ref{eq3}, where $y\textsubscript{i}$ and $\widehat{y}\textsubscript{i}$ denote the \textit{i}-th actual and predicted voltage magnitude, respectively. While the RMSE metric (\ref{eq2}) tends to penalize large prediction errors more strongly, the MAE metric (\ref{eq3}) does not weight large errors heavily compared to RMSE \cite{zhang2015suite}. Nonetheless, both metrics are easy to interpret as they are expressed in units of the variable of interest. The lower values of RMSE and MAE indicate a potentially good prediction model. 

\begin{equation}\label{eq2}
RMSE = \sqrt{\frac{1}{n}\sum_{i=1}^{n}(y\textsubscript{i} - \widehat{y}\textsubscript{i})^2}
\end{equation}
\begin{equation}\label{eq3}
MAE = \frac{1}{n}\sum_{i=1}^{n}|y\textsubscript{i} - \widehat{y}\textsubscript{i}|
\end{equation}

\subsection{Computational Setup}
Data analysis is performed in Python, including generating data sets for observability analysis and data preprocessing. Python's ML library \cite{pedregosa2011scikit} was used to implement prediction models. To ensure the reproducibility of the results, that is, to train the same prediction model on the data sets of interest, a fixed seed for the random number generator is used. The model is implemented on a computer with an Intel(R) Core(TM) i9-9900 CPU at 3.10 GHz, with 64.0 GB RAM.

\section{Numerical Results} \label{Results}
This section presents the numerical performance validation results of the proposed framework under different observability levels, separately for the LV sections of the SMART-DS feeder. The conventional ordinary least squares regression algorithm, Linear Regression (LR), is used as a benchmark method for comparison with the two tree-structured ensemble regression methods employed. The comparison was made between the results from the benchmark data set containing actual (ground-truth) voltage magnitudes, and the corresponding voltage magnitudes predicted by the learning models.

\subsection{Case Study I}

\begin{figure}[!t]
\centering
\includegraphics[width=\columnwidth]{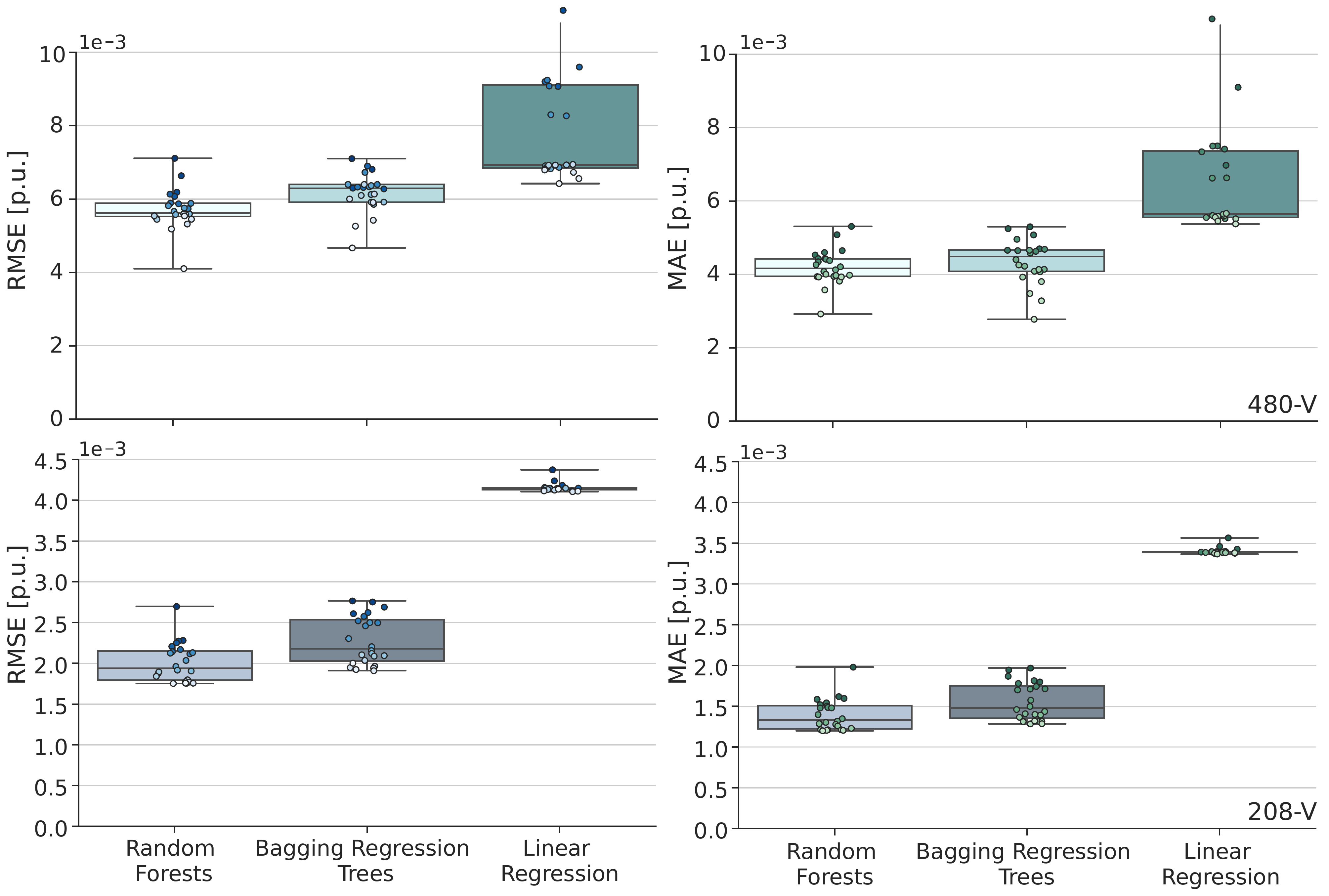}
\caption{Average RMSE and MAE of prediction models for the first case study. Color range of dots, from light to dark, represents the range of high to low-observability cases. The units are $10^{-3}$ [p.u.].}
\label{Case1_boxplots}
\end{figure}

\begin{figure}[!t]
\centering
\includegraphics[width=\columnwidth]{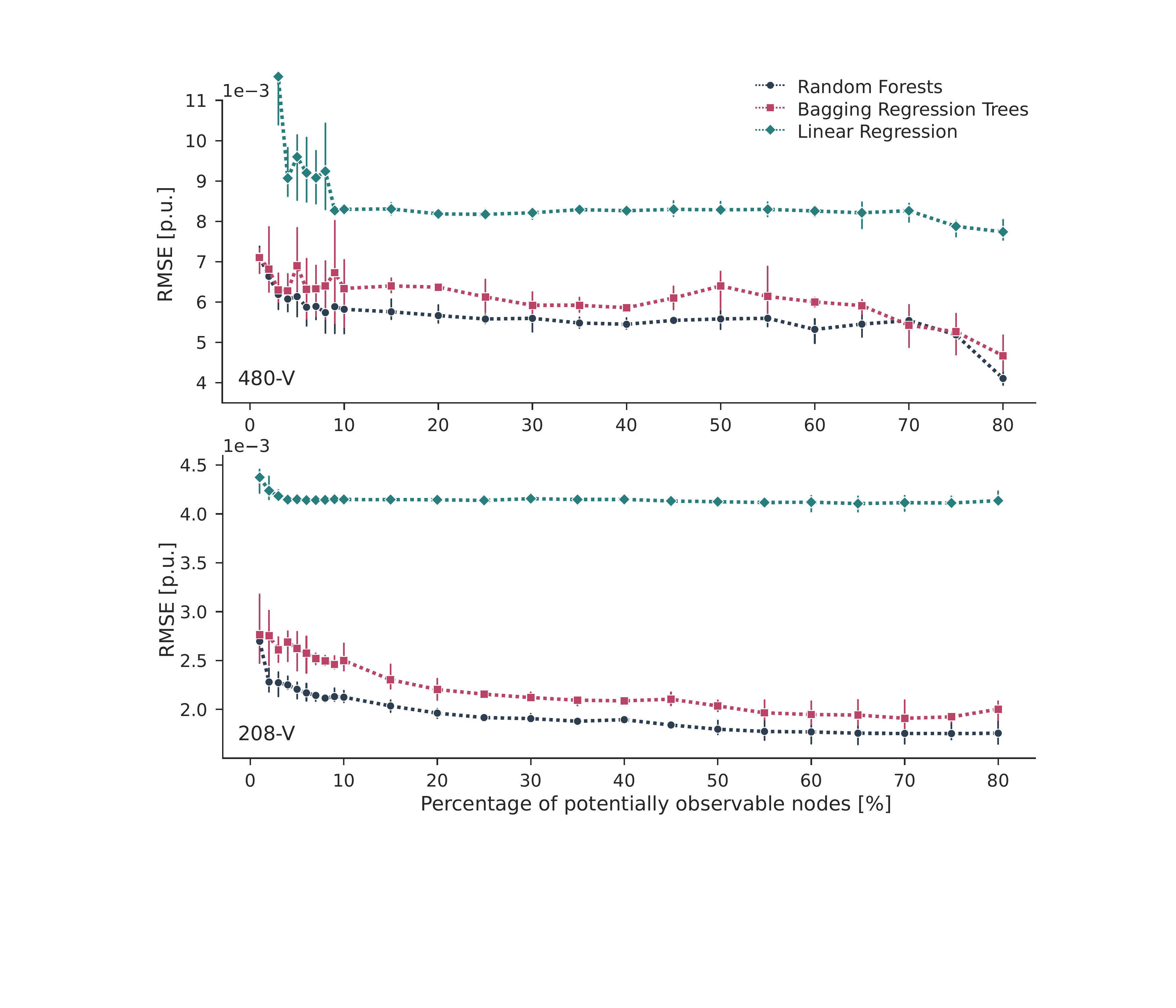}
\caption{Performance of prediction models under various observability modes for the first case study. The dashed lines represent RMSE averaged for the 3 different random samplings, whereas the discrete bars correspond to individual metrics. The units are $10^{-3}$ [p.u.].}
\label{Case1 RMSE}
\end{figure}

\begin{figure}[!t]
\centering
\includegraphics[width=\columnwidth]{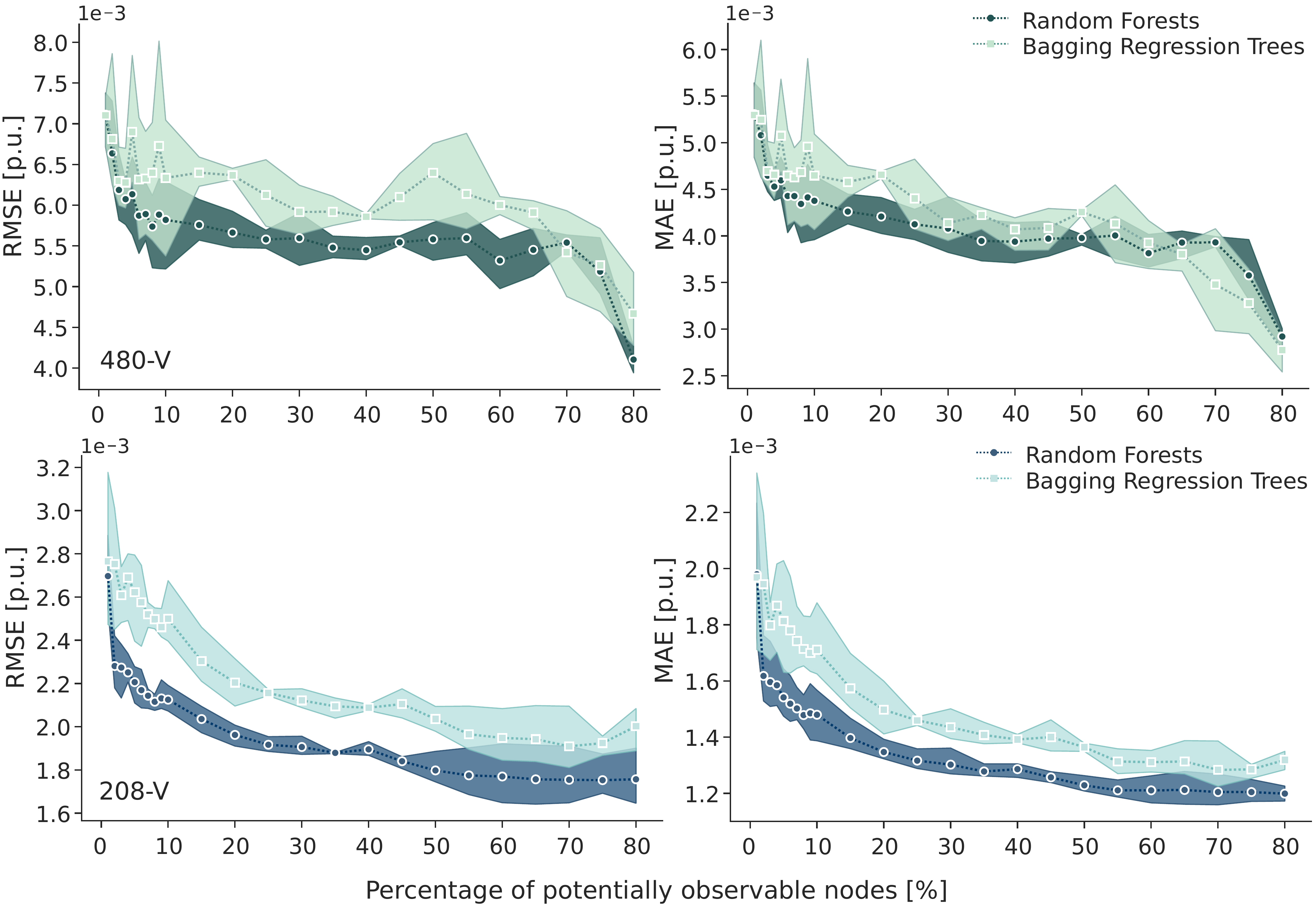}
\caption{Performance of tree-based models under various observability modes for the first case study. The dashed lines represent RMSE and MAE averaged for the 3 different random samplings, whereas the colored bands show the interval in which the corresponding individual metrics are located. The units are $10^{-3}$ [p.u.].}
\label{Case1 Observability Metrics}
\end{figure}

\begin{figure*}[!t]
	\centering
	\includegraphics[width=\textwidth,keepaspectratio]{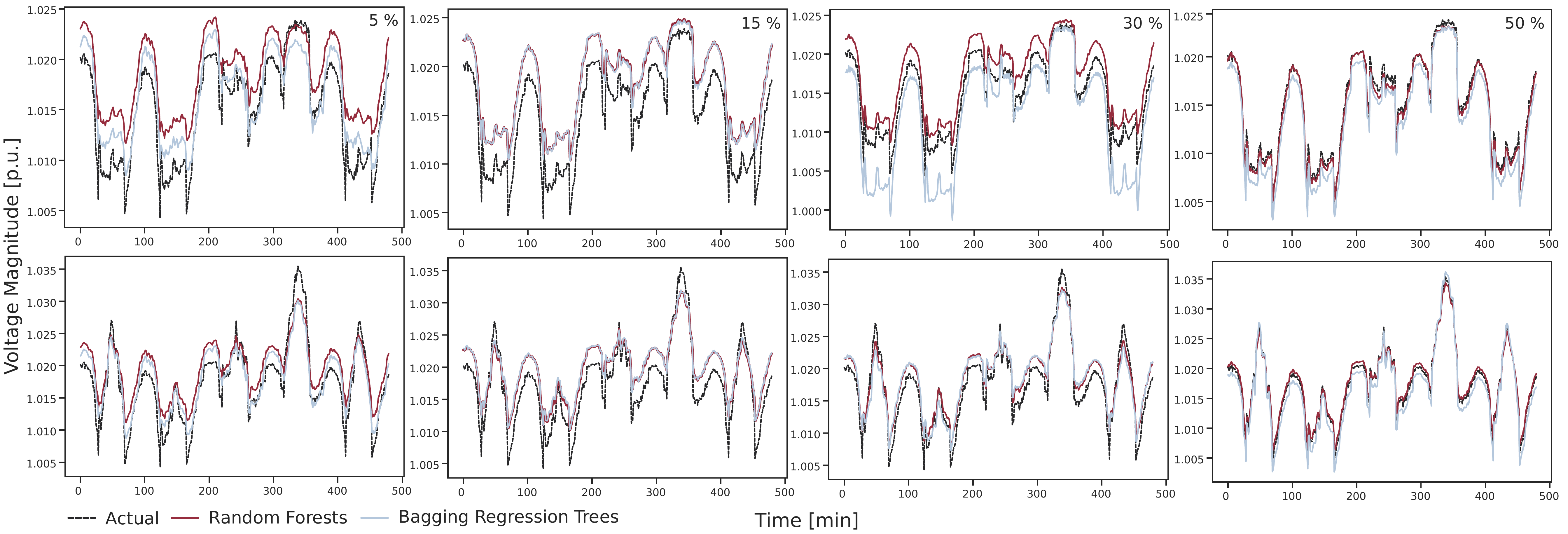}
	 \caption{\label{profile} Representative bus voltage magnitude time-series estimates in the case of the data sets with 5, 15, 30, and 50 \% of monitored nodes, respectively. Top set of figures corresponds to the case study I, while the bottom set of figures corresponds to the case study II.}
\end{figure*}

\begin{figure}[!t]
\centering
\includegraphics[width=\columnwidth]{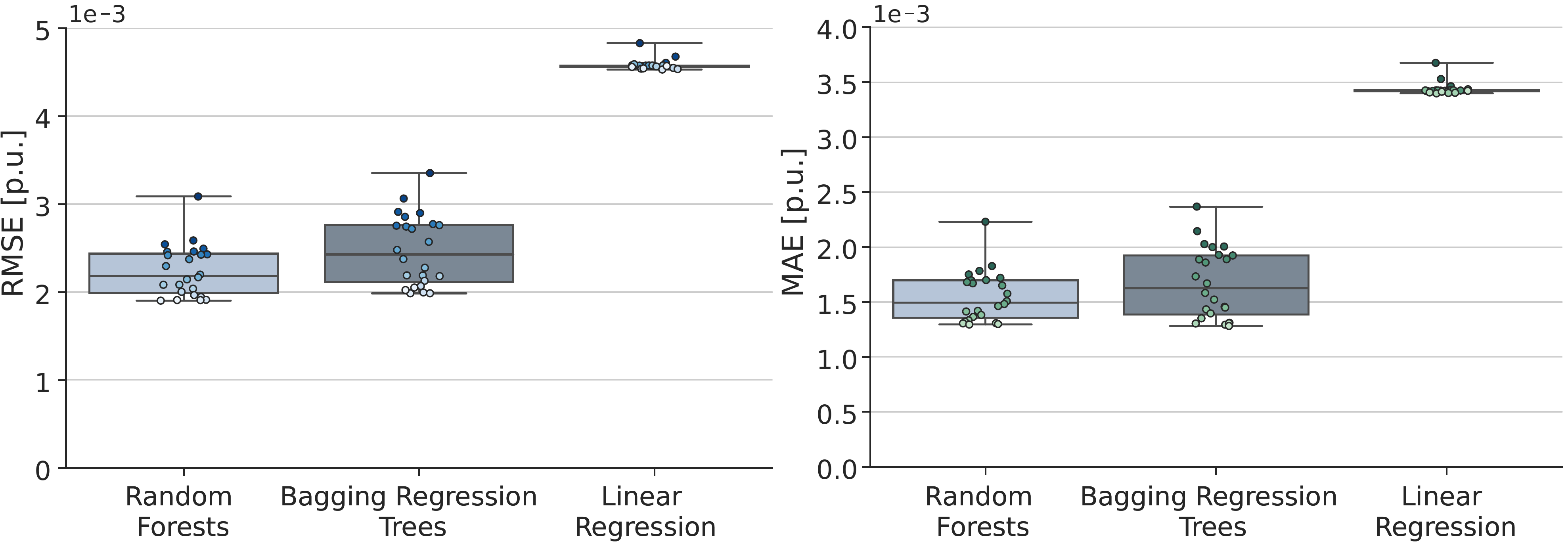}
\caption{Average RMSE and MAE of prediction models for the second case study. The units are $10^{-3}$ [p.u.].}
\label{Case2_boxplots}
\end{figure}

\begin{figure}[!t]
\centering
\includegraphics[width=\columnwidth]{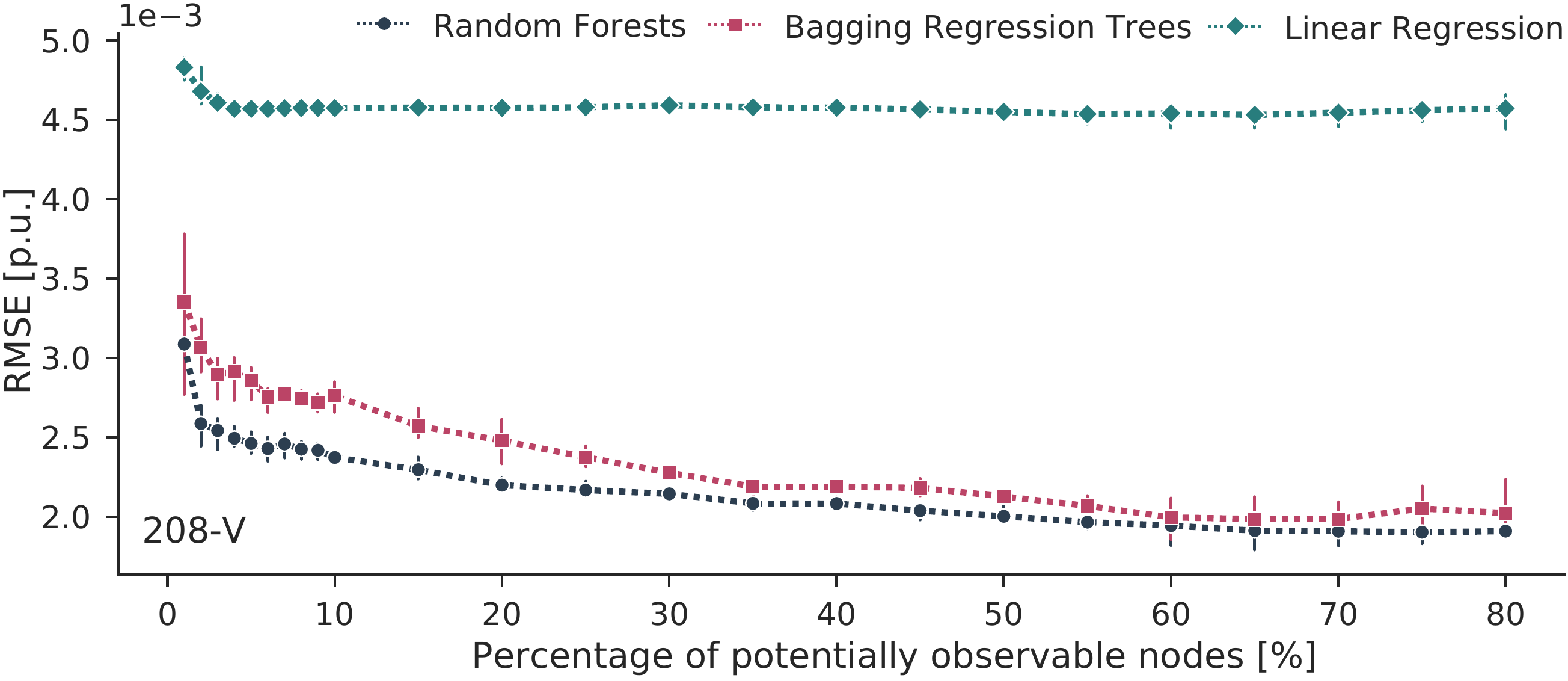}
\caption{Performance of prediction models under various observability modes for the second case study. The units are $10^{-3}$ [p.u.].}
\label{Case2 RMSE}
\end{figure}

\begin{figure}[!t]
\centering
\includegraphics[width=\columnwidth]{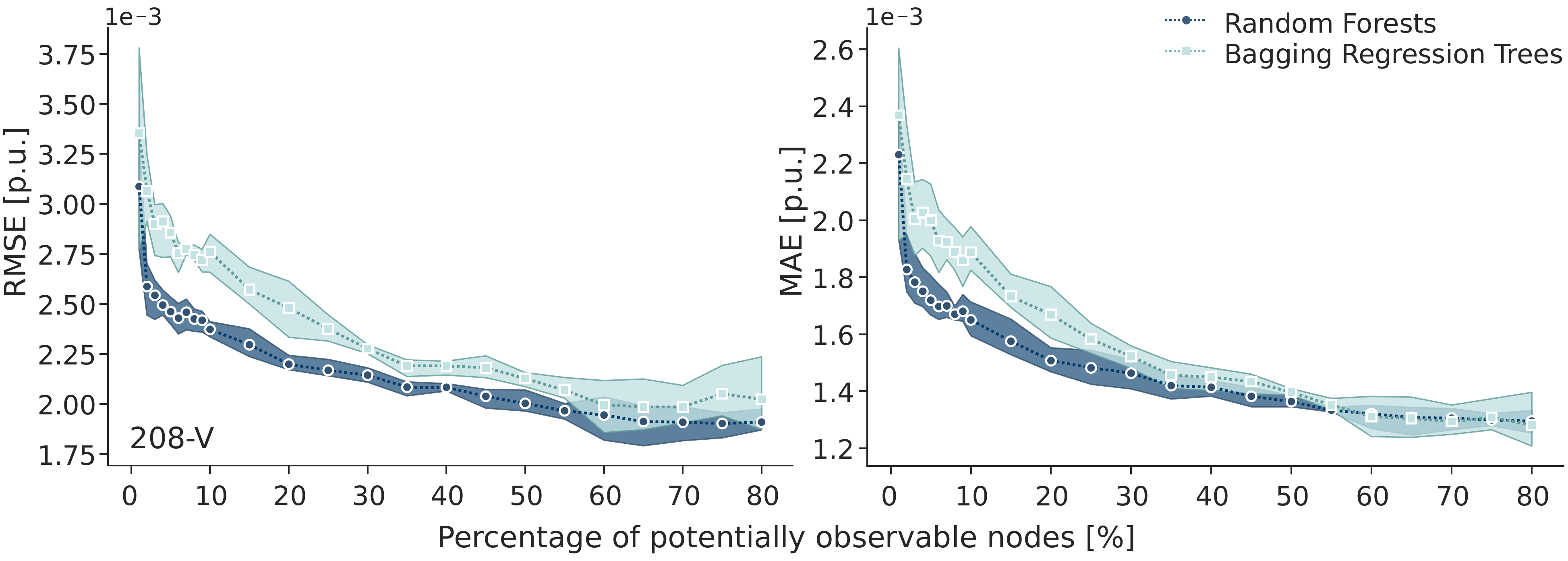}
\caption{Performance of tree-based models under various observability modes for the second case study. The units are $10^{-3}$ [p.u.].}
\label{Case2 Observability Metrics}
\end{figure}

Results of the first case study are presented for both LV sections of the examined SMART-DS feeder. On this feeder, voltage magnitudes are approximately in the range of 0.98-1.03 p.u., with more pronounced differences between voltages at 480-V. The performance indicators averaged across all observability scenarios and the random samplings are summarized in Table~\ref{tab:I}. The average RMSE/MAE of tree-based models are approximately 2.8/2.9 times higher in the case of 480-V compared to 208-V, respectively. The distribution of averaged metrics is shown in Fig.~\ref{Case1_boxplots}, where each box plot corresponds to a single regression model and graphically depicts groups of metric values through their quartiles (minimum, median, and maximum), whereas the dots refer to the metrics of the corresponding observability subcases. In Fig.~\ref{Case1 RMSE}, the averaged RMSE across all random samplings is shown for different percentages of potentially observable nodes. Recall that each analyzed observability data set is divided into training and test data sets corresponding to monitored and unmonitored nodes, respectively. Therefore, the size of the training/test sets is not the same, and the prediction models are not built on the same sized training data sets. Consequently, the metrics shown are not absolutely comparable. However, Fig.~\ref{Case1_boxplots} clearly shows the robustness of the random forest. In all analyzed observability cases, tree-based models significantly outperformed the benchmark model. When 50\% or more voltage measurements are available, the best model's RMSE drops to $1.8\cdot10^{-3}$ p.u. and below in the case of 208-V network. In the least observability cases (in which only 1-2\% of nodes are monitored) for a nominal system voltage of 480-V, the LR predictions deviate significantly from the actual values; corresponding RMSE and MAE values range from $10^5$ and $10^9$ p.u. and are, therefore, not shown in the Figs.~\ref{Case1_boxplots} -~\ref{Case1 RMSE}. Figure~\ref{Case1 Observability Metrics} depicts in more detail the evaluation metrics of the two prediction models of interest (RF and BRT) under various observability modes examined (with the percentages of monitored nodes ranging from 1\% to 80\%). It can be seen from Fig.~\ref{Case1 Observability Metrics} that RF showed consistently better estimates compared to BTR at 208-V, with very small differences between each individual random seed used. In contrast, the differences between the individual metrics are larger at 480-V, with BTR achieving slightly higher RMSE and MAE in high-observability cases.

Metric analysis alone is not sufficient to determine whether the prediction models provide satisfactory estimates in time-series predictions. For this reason, a representative node was selected and its estimated voltage magnitude is shown in Fig.~\ref{profile} assuming 5, 15, 30 and 50\% of measurements availability. The voltage magnitude estimates of RF and BRT are compared with the actual values for a time interval of 5 days, to demonstrate the scalability of the tree-based models. Both RF and BRT showed relatively high accuracy across all levels of measurement availability. On the other hand, LR (not pictured) showed inability to display voltage changes over time, thus, providing only single "snapshot" voltage estimates. When the number of monitored nodes is increased, tree-based models give more accurate estimates. This is clearly seen in Fig.~\ref{profile}; when 50\% (or more) buses are monitored, the prediction accuracy is significantly improved; moreover, the predictions are near-perfect for most buses.

\subsection{Case Study II}
As most PV systems were added at 208-V, more noticeable voltage fluctuations occur at this voltage level (see Fig.~\ref{Voltages Distribution}). For this reason, the results of two case studies for a nominal system voltage of 480-V exhibit a similar behavior; thus, only results for the nominal system voltage of 208-V level are reported here. The performance indicators averaged across all observability scenarios and the random samplings are summarized in Table~\ref{tab:II}. The average RMSE/MAE of tree-based models are approximately $2\cdot10^{-4}$/ $0.13\cdot10^{-4}$ p.u. higher in the second case study compared to first case study, respectively. As in the previous case study, the distribution of averaged metrics is shown in Fig.~\ref{Case2_boxplots}, whereas the averaged RMSE across all random samplings is shown for different percentages of potentially observable nodes in Fig.~\ref{Case2 RMSE}. Similarly, the tree-based methods outperformed the benchmark model, with RF achieving better voltage estimates on average compared to BRT. Figure~\ref{Case2 Observability Metrics} depicts in more detail the evaluation metrics of the two prediction models of interest (RF and BRT) under various observability modes examined (with the percentages of monitored nodes ranging from 1\% to 80\%). It can be seen from Fig.~\ref{Case2 Observability Metrics} that RF showed consistently better RMSE compared to BTR under all observability ranges, whereas in moderate-observability cases (when 60\% and 65\% of nodes are monitored), BTR achieved only slightly higher MAE. 

Time-series performance of the two models is shown at the bottom of the Fig.~\ref{profile}, where the same bus is selected for comparison. Analogous to the first case study, when the number of monitored nodes is increased, tree-based models give more accurate estimates. Based on the results presented here, the prediction model demonstrates robustness to voltage rise due to the presence of PV systems in the network, as long as the prediction model is trained on the available voltage measurements that reflect current network operating conditions. 

\section{Conclusion} \label{Conclusion}

This paper proposed the use of the existing CATV communications infrastructure to leverage voltage measurement capabilities of sensors deployed in CATV networks for voltage estimation in LV distribution grids. The presented framework based on random forest estimates the voltage magnitudes at non-monitored buses in LV distribution grids using this timely CATV sensing. The prospects of accurate LV estimation even under low-observability conditions were demonstrated using a distribution feeder from the SMART-DS data set, with and without distributed PV. The results show that voltage estimates are robust and insensitive to the low percentages of monitored nodes. 

Further research is needed to support the idea behind the proposed use of CATV measurements. In particular, one potential drawback is that the CATV sensor locations are physically and electrically decoupled from the distribution grid nodes. However, since CATV sensors are in the immediate vicinity of electric grid nodes, cluster analysis can address the disputed mismatch. For example, an algorithmic clustering solution that yields weighted cluster rankings according to distance-based similarity can be used to match a CATV sensor to adjacent LV grid node. Additionally, more accurate voltage estimates could be achieved by splitting the learning model into several local regressors, depending on the size of the feeder and the variability of the voltage profile distributed across the feeder. The proposed approach does not expect an increase in computational burden due to the possibility of parallel computing.

\section*{Acknowledgment}
A special thanks to Josh Bauer with the National Renewable Energy Laboratory (NREL) graphics team for the creation of Fig.~\ref{Connected Neighborhood}.
\ifCLASSOPTIONcaptionsoff
  \newpage
\fi



\bibliographystyle{IEEEtran}
\end{document}